\documentclass[prd,aps,onecolumn,nofootinbib,notitlepage]{revtex4}

\usepackage{enumitem}
\usepackage{graphicx}
\usepackage{subfigure}
\usepackage{amssymb}
\usepackage{amsmath}
\usepackage{bbm}
\usepackage{color}
\usepackage[retainorgcmds]{IEEEtrantools}

\newcommand{\huts}{Hutsem\'{e}kers }
\newcommand{\tab}{~~}
\newcommand{\beq}{\begin{equation}}
\newcommand{\eeq}{\end{equation}}
\newcommand{\ignore}[1]{}
\newcommand{\be}{\begin{equation}} \newcommand{\ee}{\end{equation}}
\newcommand{\bea}{\begin{eqnarray}} \newcommand{\eea}{\end{eqnarray}}
 \renewcommand{\bf}{\textbf}

\begin{document}

\title{A complete $3D$ numerical study of the effects of pseudoscalar-photon mixing on quasar polarizations}

\author{Nishant Agarwal$^1$}
\author{Pavan K. Aluri$^2$}
\author{Pankaj Jain$^2$}
\author{Udit Khanna$^2$\footnote{Present address: Harish-Chandra Research Institute, Allahabad - 211019, India}}
\author{Prabhakar Tiwari$^2$}

\affiliation{
$^1$McWilliams Center for Cosmology, Department of Physics, Carnegie Mellon 
University, Pittsburgh, PA - 15213, USA \\ 
$^2$Department of Physics, Indian Institute of Technology, Kanpur - 208016, India\\
}

\begin{abstract}
We present the results of three-dimensional simulations of quasar polarizations in the presence of pseudoscalar-photon mixing in the intergalactic medium. 
The intergalactic magnetic field is assumed to be  
uncorrelated in wave vector space but correlated in real space. 
Such a field may be obtained if its origin is primordial.
 Furthermore we assume that the quasars,
located at cosmological distances, have negligible initial polarization. 
In the presence of pseudoscalar-photon mixing we show, through a direct comparison with observations, that this may explain the observed large scale alignments in quasar polarizations within the framework of big bang cosmology. 
We find that the simulation results give a reasonably good fit to the observed 
data.
\end{abstract}

\maketitle


\section{Introduction}
\label{one}
In a recent paper \cite{Agarwal:2011} 
it was shown that pseudoscalar-photon mixing in the presence of 
correlated intergalactic magnetic fields might provide an explanation for
the observed large scale alignment of quasar polarizations at visible 
wavelengths \cite{Hutsemekers:1998,Hutsemekers:2000fv,Hutsemekers:2005iz}. 
The alignment is seen over cosmologically large distances of order Gpc
\cite{Hutsemekers:1998,Hutsemekers:2000fv,Jain:2002vx,Jain:2003sg,Hutsemekers:2005iz,Payez:2008pm,Piotrovich:2008iy}. 
In \cite{Hutsemekers:2005iz}, the authors argue that it is
very unlikely that the effect originates due to interstellar extinction.
The effect is puzzling within the framework of big bang cosmology since 
we do not expect astrophysical objects to be correlated over such large
distances. There is also direct evidence that other properties of these
objects, such as, polarization position angles at radio frequencies or
position angles of parsec-scale jets, do not show any large scale alignment
\cite{Joshi:2007}.
Hence the effect does not seem to be intrinsic to these sources and might
arise due to propagation. Furthermore the propagation effect must not affect radio
polarizations since these do not show any alignment. Pseudoscalar-photon mixing is a good candidate to explain the alignment of quasar optical polarizations since it is negligible at radio frequencies. Large scale correlations in the intergalactic magnetic field could lead to such an alignment over large distances \cite{Agarwal:2011}. 

In \cite{Agarwal:2011}, it was assumed that the intergalactic
magnetic field may be obtained by a simple cosmological evolution of
the primordial magnetic field
\cite{Subramanian:2003sh,Seshadri:2005aa,Seshadri:2009sy,Jedamzik:1996wp,Subramanian:1997gi} within the framework of big bang cosmology. It was assumed that visible radiation
from distant quasars is unpolarized at the source. It acquires  
polarization due to its mixing with hypothetical low mass
pseudoscalars in background magnetic fields 
\cite{Clarke:1982,Sikivie:1983ip,Sikivie:1985yu,Sikivie:1988mz,Maiani:1986md,Raffelt:1987im,Carlson:1994yqa,Bradley:2003kg,Das:2004qka,Das:2004ee,Ganguly:2005se,Ganguly:2008kh}. Such pseudoscalars arise in many extensions of the standard model of particle
physics \cite{Peccei:1977hh,Peccei:1977ur,Weinberg:1977ma,Wilczek:1977pj,McKay:1977gd,McKay:1978wn,Kim:1979if,Dine:1981rt,Kim:1986ax}. It was shown that the linear 
polarization angle of electromagnetic radiation from quasars separated over 
large distances becomes aligned due to correlations in the background
magnetic field \cite{Agarwal:2011}. In \cite{Agarwal:2011} 
the authors restricted their attention to quasars that
are aligned in one dimension, along the line of sight. In the current paper
we present the results of a general three-dimensional numerical analysis 
of this problem, and show by a direct comparison with observations that such an effect can explain the observed large scale alignments in quasar polarizations.

The mixing of photons with pseudoscalars has many interesting astrophysical and 
cosmological implications \cite{Harari:1992ea,Berezhiani,Mohanty:1993nh,Das:2000ph,Kar:2000ct,Kar:2001eb,Csaki:2001jk,Csaki:2001yk,Grossman:2002by,Jain:2002vx,Song:2005af,Mirizzi:2005ng,Raffelt:2006cw,Gnedin:2006fq,Mirizzi:2007hr,Finelli:2008jv,DeAngelis:2007,DeAngelis:2009,DeAngelis:2011}. It affects both the intensity and polarization of radiation. Furthermore
it also changes its spectral characteristics \cite{Ostman:2004eh,Lai:2006af,Hooper:2007bq,Hochmuth:2007hk,Chelouche:2008ta}. Extensive laboratory and 
astrophysical searches of these particles have led to stringent limits on 
their masses and couplings 
\cite{Dicus:1978fp,Vysotsskii:1978,Dearborn:1985gp,Raffelt:1987yu,Raffelt:1987yt,Turner:1987by,Mohanty:1993nh,Janka:1995ir,Keil:1996ju,Brockway:1996yr,Grifols:1996id,Raffelt:1999tx,Rosenberg:2000wb,Zioutas:2004hi,Yao:2006px,Raffelt:2006cw,Jaeckel:2006xm,Andriamonje:2007ew,Robilliard:2007bq,Zavattini:2007ee,Rubbia:2007hf,Lee:2006za,Agarwal:2008ac}. Some astrophysical observations also 
appear to indicate a positive signature of pseudoscalar-photon mixing
\cite{DeAngelis:2007,Bassan:2010}  

The intergalactic magnetic field is not known very well. In most treatments
of pseudoscalar-photon mixing through intergalactic space it is assumed 
that the background medium may be split into a large number of uncorrelated
domains. The domain size is typically taken to be of order Mpc with the
magnetic field in each domain of order nG 
\cite{Csaki:2001yk,Grossman:2002by,Mirizzi:2005ng}. 
The magnetic field in different domains points in random
directions and is uncorrelated. Here we  
assume that the background magnetic field is of primordial origin
\cite{Subramanian:2003sh,Seshadri:2005aa,Seshadri:2009sy,Jedamzik:1996wp,Subramanian:1997gi}. We also consider discretized real space consisting of a large number of cells or domains
of equal size. Each domain is assumed to be cubical of side $r_G$. 
The $j^{\rm th}$ component of the magnetic field, $B_j({\boldsymbol r})$, 
in real space is given by the
discrete Fourier transform, 
\bea
B_{j}({\boldsymbol r}) = \frac{1}{V} \sum b_{j}({\boldsymbol k}) e^{i {\boldsymbol k}.{\boldsymbol r}},
\label{eq:IFT}
\eea
where $V$ is the volume in real space and 
$b_j({\boldsymbol k})$ is the $j^{\rm th}$ component of the  
magnetic field in wave vector space. The magnetic field is assumed to be uniform
in each domain. We may write the two point correlations of $b_j({\boldsymbol k})$ as, 
\bea
	\langle b_{i}({\boldsymbol k}) b^{*}_{j} ({\boldsymbol q}) \rangle & = & \delta_{{\boldsymbol{k,q}}} P_{ij}({\boldsymbol k}) M(k) \nonumber \\
	& = & \delta_{{\boldsymbol{k,q}}} \sigma^{2}_{ij}({\boldsymbol k}),
\label{eq:corr}
\eea
where $\sigma^{2}_{ij}({\boldsymbol k}) = P_{ij}({\boldsymbol k}) M(k)$, 
$k = |{\boldsymbol k}|$ and 
$P_{ij}({\boldsymbol k}) = \left(\delta_{ij} -\frac{k_{i} k_{j}}{k^2}\right)$ is the projection operator. The function $M(k)$ shows a power law behaviour,
\be
    M(k) = A k^{n_{B}}\,,
    \label{eqn2}
\ee
where $n_{B}$ is the power spectral index. The power law dependence of this correlation
function would lead to correlations in the magnetic field in real space over large
distances. This is in contrast to the assumption made in most treatments
of pseudoscalar-photon mixing through intergalactic space
\cite{Csaki:2001yk,Grossman:2002by,Mirizzi:2005ng}. 
 The constant $A$ in Eq. \ref{eqn2} is fixed by demanding that 
\be
\sum_i <B_i({\boldsymbol r}) B_i({\boldsymbol r})> = B_0^2,
\ee
where we assume the value of about 
1 nG \cite{Seshadri:2009sy,Yamazaki:2010nf,DeAngelis:2008} for $B_0$.

The value of the spectral index $n_B$
has been obtained by making a best fit to 
the matter and CMB power spectrum \cite{Yamazaki:2010nf}. This fit 
gives, $n_{B} = -2.37$. A fully scale invariant power spectrum would
correspond to the value $n_{B} = -3$. We expect that the  
intergalactic medium may be turbulent on short distance scales. 
In this case we may expect the 
magnetic field to obey a Kolmogorov power spectrum for distance scales
much smaller than the size of the system. A similar model is also applicable
to the galactic medium, where both the plasma density and the magnetic
field are known to show a Kolmogorov power 
spectrum \cite{Armstrong:1995,Minter:1996}. Hence it might be more 
appropriate to assume that $n_B$ has some dependence on $k$. 
Here we shall simply treat $n_B$ as a fixed parameter, independent of $k$. 
The spectrum may also have a 
low $k$ cutoff, $k_{\rm min}$, which will imply a cutoff
on correlations in real space for distances larger than $r_{\rm max}=k^{-1}_{\rm min}$.
Here we shall assume that $r_{\rm max}$ is larger than the size of our 
system, which we take to be of the order of a few Gpc. Hence we may simply set it equal 
to infinity. The large scale correlations induced among quasar polarizations 
crucially depend on this parameter. If this parameter is much smaller than
1 Gpc, then the mechanism proposed in \cite{Agarwal:2011} may not work. 
Alternatively we may argue that the observation of large scale correlations
in quasar polarizations might be an indication that $r_{\rm max}\ge 1$ Gpc. 
We emphasize that a magnetic field with such large distance 
correlations may be generated 
 within the framework of the big bang model. The perturbations
generated at the time of inflation have wavelengths extending up to the
horizon in the current era. The Cosmic Microwave Background (CMB) 
anisotropies show significant power in the quadrupole, which implies 
correlations over the entire sky. The matter spectrum, however, gets 
significantly modified due to evolution after decoupling and correlations
at such large distances get suppressed. However, the magnetic field need
not evolve in the same manner as the non-relativistic matter.  

In our analysis we first ignore cosmological evolution and assume
a flat space-time. We are interested in radiation from quasars located
at redshifts of order unity. In this case our neglect of cosmological
evolution would lead to an error of order unity, which is comparable to
other errors, such as in the modelling of the intergalactic magnetic field.
So far most of the literature on pseudoscalar-photon mixing in 
a background magnetic field has ignored cosmological evolution. 
We next also take cosmological evolution into account. 

A useful statistic to test for alignment of quasar polarizations is defined as
follows \cite{Hutsemekers:1998,Jain:2003sg}. We first define a measure of dispersion, $d_k$, in the neighbourhood
of the $k^{th}$ quasar, 
\bea
d_{k} = \frac{1}{n_{v}} \sum_{i=1}^{n_{v}} \cos[2(\psi_{i}+\Delta_{i\rightarrow
k}) - 2\psi_{k})].
\label{eq:dispersion}
\eea
Here $\psi_i$ are the polarization angles of the nearest neighbours of the 
quasar at position $k$. We 
include a total of $n_v$ nearest neighbours. The nearest neighbours also
include the polarization angle of the $k^{\rm th}$ quasar. 
The concept of nearest neighbours requires a measure of distance. 
In a curved space we do not have a unique definition of distance. Here
we use comoving distance as well as angular diameter distance for this
purpose. When we include cosmological evolution, we avoid this problem
by locating our simulated sources at the precise positions of the observed
sources. 
The polarizations at two different angular positions 
are compared by making a parallel transport from their location to the 
position of the $k^{\rm th}$ quasar along the great circle joining these points
on the celestial sphere. This induces
the factor $\Delta_{i\rightarrow k}$, included in the definition of $d_k$.
The parallel transport is required since we are comparing polarizations
located at different points on the surface of the celestial sphere.
If this parallel transport factor is not included then the resulting 
statistic is not invariant under coordinate transformations \cite{Jain:2003sg}. 
We next maximize $d_k$ as a function of $\psi_k$. The resulting value of
$\psi_k$ is interpreted as the mean polarization angle at the position $k$ and
the corresponding maxima of the function in Eq. \ref{eq:dispersion} gives
an estimate of $d_k$. 
 The statistic may now be defined as \cite{Hutsemekers:1998,Jain:2003sg},
\bea
	S_{D} = \frac{1}{n_{s}} \sum_{k=1}^{n_{s}} d_{k}{\big |}_{\rm max},
\label{eq:statistic}
\eea
where $n_s$ is the total number of sources in the data. 
A large value of $S_{D}$ indicates a strong alignment between polarization vectors. In our analysis we shall compute this statistic theoretically
and compare the corresponding values 
obtained by observations \cite{Hutsemekers:1998,Hutsemekers:2000fv,Hutsemekers:2005iz, Jain:2003sg} in order get an estimate of the model parameters. 

A potential problem with our hypothesis that the alignment is caused by
pseudoscalar-photon mixing is discussed in Ref. \cite{Hutsemekers:2011}. 
It is observed that in most cases these quasars show 
negligible circular polarization
\cite{Hutsemekers:2005iz,Hutsemekers:2010}. So far this has been seen only
in a small sample. Pseudoscalar-photon mixing predicts a larger 
circular polarization and hence may not consistently explain the data. 
The authors argue that if we assume the incident light to be natural 
white light, then due to decoherence the circular polarization is 
predicted to be zero, consistent with observations. However, circular polarization is observationally found to be very small even with a broadband filter. 
In this case theory does indicate a significant circular polarization which
is not in agreement with the data. Here we address this issue partially. 
We point out that there exist many other effects which may cause decoherence
and hence reduce the predicted circular polarization. For example, 
in most treatments the medium is assumed to be uniform. Alternatively
one assumes a large number of domains with different properties but
each individual domain is assumed to be uniform. In reality 
the medium, at any reasonable length scale, 
shows fluctuations both in space and time. Since the background
magnetic field is not uniform, the pseudoscalar particle produced would not
have the same energy and momentum as that of the incident photon. This
acts as another source of decoherence and will 
suppress the circular polarization. 
We show this explicitly by considering a model space varying
magnetic field. We find that if the magnetic field shows rapid variations
in space, then the circular polarization is greatly reduced in comparison
to the linear polarization. Furthermore we argue that a time varying medium
may also lead to suppression of circular polarization.

\section{Pseudoscalar-photon mixing in an expanding universe}

In this section we give the basic equations of pseudoscalar-photon mixing
in an expanding universe \cite{Carroll:1991,Garretson:1992,Cudell:2011}. We assume the standard 
Friedmann-Robertson-Walker (FRW) metric, with the curvature parameter $k=0$, 
using conformal time.  
The metric may be written as,
\begin{equation}
{g_{\mu\nu}} = \left( \begin{array}{cccc} 
a^{2} & 0 & 0 & 0  \\ 
0 & -a^{2} & 0 & 0 \\ 
0 & 0 & -a^{2} & 0 \\ 
0 & 0 & 0 & -a^{2} \\ 
\end{array} \right) = a^{2}\eta_{\mu\nu},
\end{equation}
where $a$ is the scale parameter and 
$\eta_{\mu\nu}$ the Minkowski metric. 


\subsection{Mixing with pseudoscalars}

The action of the electromagnetic field $F_{\mu\nu}$ coupled to a pseudoscalar
 field $\phi$ may be written as,
\begin{IEEEeqnarray}{rCl} 
S= \int d^{4}x \sqrt{-g}~ \Big[-\frac{1}{4}F_{\mu\nu}F^{\mu\nu}-\frac{1}{4} g_{\phi} \phi F_{\mu\nu}\tilde{F}^{\mu\nu}
\nonumber\\+\frac{1}{2}(\omega_{p}^{2}a^{-3})A_{\mu}A^{\mu}+\frac{1}{2}g^{\mu\nu}\phi_{,\mu}\phi_{,\nu}-
\frac{1}{2}m^{2}_{\phi}\phi^{2}\Big].
\label{eq:S_expanding}
\end{IEEEeqnarray}
Here $g_{\phi}$ is the 
coupling of $\phi$ with the electromagnetic field. We have also written
 an effective photon mass term. This corresponds to the effective mass 
$\omega_{p}^{2}a^{-3}$ it
acquires due to the medium, where $\omega_p$ is the plasma frequency.
In Eq. \ref{eq:S_expanding} 
we have ignored any self couplings of the scalar field.
The equation of motion for the scalar field may be written as,
\beq 
-\frac{\partial^{2}\chi}{\partial \eta^{2}} + \nabla^{2}\chi= -\frac{g_{\phi}}{a}(a^{2}{\boldsymbol E})\cdot (a^{2}{\boldsymbol {\mathcal B}})+ m^{2}_{\phi} a^{2}\chi ,
\label{eq:EOM_exp}
\eeq
where we have replaced $\phi$ by $\frac{\chi}{a}$. In Eq. \ref{eq:EOM_exp},
${\mathcal{B}_{i}}+B_{i}=\frac{1}{2}\epsilon_{ijk}a^{2}F^{jk}$ is the magnetic
field, $E_{i}=a^{2}F^{0i}$ the usual electric field, ${\boldsymbol {\mathcal B}}$ the
background magnetic field, and ${\boldsymbol B}$ the magnetic field of the 
wave. The Maxwell's equations in curved space-time 
may be written as,
\beq {\boldsymbol \nabla}.(a^{2}{\boldsymbol E}) = -\frac{g_{\phi}}{a}{\boldsymbol \nabla}\chi . (a^{2}{\boldsymbol {\mathcal B}})+\left(\frac{\omega_{p}^{2}}{a}\right)A_{0},\eeq 
\beq {\boldsymbol \nabla} \times (a^{2} {\boldsymbol E}) +\frac{\partial{(a^{2}{\boldsymbol B})}}{\partial\eta} = 0,\eeq
with 
\beq \frac{\partial{(a^{2}{\boldsymbol {\mathcal B}})}}{\partial\eta} = 0,\eeq
\beq {\boldsymbol \nabla} \times (a^{2}{\boldsymbol B}) -\frac{\partial{(a^{2}{\boldsymbol E})}}{\partial \eta} = 
     \frac{g_{\phi}}{a}({\boldsymbol \nabla}\chi\times a^2{\boldsymbol E})+\frac{g_{\phi}}{a}(a^{2}{\boldsymbol {\mathcal B}}) \frac{\partial\chi}{\partial \eta}+\left(\frac{\omega_{p}^{2}}{a}\right){\boldsymbol A},\eeq
\beq  {\boldsymbol \nabla}.(a^{2}{\boldsymbol B})=0, \eeq
where ${\boldsymbol A}$ represents the vector potential. 
In the above equations we have assumed a uniform background magnetic 
field and approximated $\boldsymbol{\mathcal{B}}+ {\bf B}\approx
\boldsymbol{\mathcal{B}}$ as 
$|{\boldsymbol {\mathcal B}}|\gg |{\boldsymbol B}|$. Here we also ignore 
the term containing derivatives of the scale factor, $a$, since the 
time scale for cosmological evolution is much larger in comparison
to other times scales in the problem.
Now we take the curl of Faraday's Law, and use the above equations to get the  wave equation for 
${\boldsymbol E}$,
\beq -\frac{\partial^{2}(a^{2}{\boldsymbol E})}{\partial \eta^{2}} + \nabla^{2}(a^{2}{\boldsymbol E})= 
\frac{\omega_{p}^{2}}{a}(a^{2}{\boldsymbol E})+\frac{g_{\phi}}{a}(a^{2}{\boldsymbol {\mathcal B}})\frac{\partial^{2}\chi}{\partial \eta^{2}}. \eeq
The time dependence of the resulting solution may be expressed as,
\beq (a^{2}E) \propto {\rm exp} (-i\omega\eta)= {\rm exp}\left[-i\omega \int \frac{{\rm d}t}{a(t)}\right].\eeq
We also note that $\omega(t)a(t)= \omega_0$, where $\omega_0$ is the frequency
at the current time.

\subsection{Mixing solution}
\label{sc:solution}
We choose the comoving coordinate system with the $z$-axis as the direction of propagation. In a particular domain the magnetic field is assumed to be
uniform. Only the component of ${\boldsymbol E}$ parallel to the background magnetic 
field mixes with $\chi$. We also define 
$ {\boldsymbol {\cal A}}= \frac{(a^{2}{\boldsymbol E})}{\omega}$ and replace $(a^{2}{\boldsymbol E})$ by $\omega {\boldsymbol {\cal A}}$. We write the field equations of
${\cal A}_{\parallel}$ and $\chi$ as,
\beq 
(\omega^{2}+ \partial^{2}_{z})\left(\begin{array}{c}
{\cal A}_{\parallel}\\ \chi \end{array} \right) - 
M \left( \begin{array}{c} {\cal A}_{\parallel}\\ \chi \end{array} \right) =0, \eeq
where,
\beq M = \left(\begin{array}{cc} 
         \frac{\omega^{2}_{p}}{a}     & ~~~ -\frac{g_{\phi}}{a}(a^{2}
\mathcal{B}_\perp)\omega\\
-\frac{g_{\phi}}{a}\ (a^{2}\mathcal{B}_\perp)\omega   & ~~~ m^{2}_{\phi}a^{2} \end{array}\right). \eeq
Here, $\mathcal{B}_\perp$ is the transverse component of ${\boldsymbol {\mathcal B}}$
and $\omega$ is the frequency of radiation at 
the propagation domain. Due to expansion the observed  energy 
today is redshifted. Hence we replace 
$\omega\rightarrow \frac{\omega}{a}$. Finally the
mixing matrix may be written as, 
\beq M = \left(\begin{array}{cc}
      
         \frac{\omega^{2}_{p}}{a}     & ~~~ -\frac{g_{\phi}}{a^{2}}
(a^{2}\mathcal{B}_\perp)\omega\\
-\frac{g_{\phi}}{a^{2}}\ (a^{2}\mathcal{B}_\perp)\omega   & ~~~ m^{2}_{\phi}a^{2} \end{array}\right). \eeq

We solve the mixed field equations in a manner similar to that in 
Ref. \cite{Das:2004qka,Das:2004ee}. We diagonalize 
the mixing matrix $M$ by an orthogonal transformation, $OMO^{T}=M_{D}$, where,
 \beq O = \left(\begin{array}{cc}
         \cos\theta  &~~~  -\sin\theta\\
         \sin\theta  & ~~~  \cos\theta \end{array} \right), \eeq
 where the angle $\theta$ can be expressed as,
  \beq    \tan2\theta = \frac{2g_{\phi}\omega a^{-2} (a^{2}\mathcal{B})}{\left(\frac{\omega^{2}_{p}}{a}-m^{2}_{\phi}a^{2}\right)}. \eeq
We assume that the  mass of the pseudoscalar ($m_{\phi}$) is negligible compared to the plasma frequency ($\omega_{p}$).

\section{Simulations}
\label{sc:simulations}
We first generate the magnetic field numerically in wave vector space. 
This is relatively straightforward since it is uncorrelated for different 
wave vectors. The projection operator implies that the component of the
magnetic field parallel to ${\boldsymbol k}$ is zero. The two orthogonal components
are uncorrelated. It is simplest to use polar coordinates $(k,\theta,
\phi)$ in wave vector space. Eq. \ref{eq:corr} implies that for any
wave vector ${\boldsymbol k}$, $b_k=0$ and $b_\theta$ and $b_\phi$ are uncorrelated.
We may, therefore, generate these by assuming the Gaussian distribution, 
\begin{eqnarray}
	f(b_{\theta}({\boldsymbol k}),b_{\phi}({\boldsymbol k})) = N \ {\rm exp} 
\left[-\left(\frac{b_{\theta}^{2}({\boldsymbol k}) + b_{\phi}^{2}({\boldsymbol k})}{2M({\boldsymbol k})}\right) \right],
\end{eqnarray}
where $N$ is the normalization factor. This represents an uncorrelated Gaussian distribution for the two components of the magnetic field in Fourier space, 
corresponding to the wave vector ${\boldsymbol k}$. Once we have these two components
we can obtain the three Cartesian components of the magnetic field in 
wave vector space. Next we use Eq. \ref{eq:IFT} to obtain the three 
Cartesian components of the magnetic field in real space.  
The Gaussian random variates are generated by using the Numerical 
Recipes \cite{Recipes:1992} code {\it gasdev}. The discrete Fourier transform is obtained by
using the Recipes code {\it fourn}. 

The optical polarizations are propagated, taking pseudoscalar-photon
mixing into account by the procedure described in Ref. \cite{Agarwal:2008ac}.
We need to propagate over a large number of magnetic domains from the
quasar to the observer. We choose a fixed ``external" coordinate system. 
The transverse component of the magnetic field in each domain is aligned
at some angle to the fixed coordinate axes. In order to propagate 
through each domain we first rotate the coordinates so that the 
transverse magnetic field aligns along one of the coordinate axes. We then
use standard expressions for pseudoscalar-photon mixing \cite{Agarwal:2008ac} in order to 
evaluate the correlation functions of the electromagnetic and pseudoscalar
fields after propagation through the domain. We then rotate back to the fixed
coordinate system. This procedure is repeated for propagation through
each domain till the wave reaches the observer.



\begin{figure}[!t]
  \centering{
    \includegraphics[width=4.5in,angle=0]{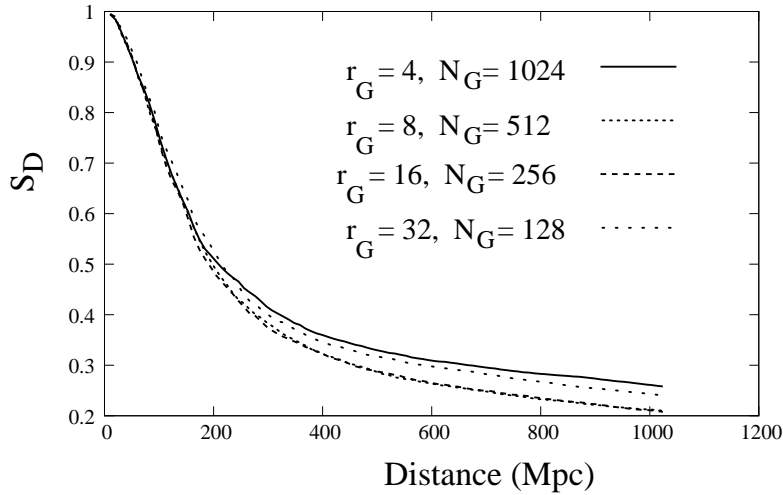}
    \caption{The statistic $S_D$ as a function of the distance between
sources for different choices of the domain size $r_G$ (in Mpc) 
and the number of
points on the grid $N_G$.  
 Here we have set $n_B=-2.37$. 
}   
\label{fig:Sd}}
\end{figure}

\begin{figure}[!t]
  \centering{
    \includegraphics[width=4.5in,angle=0]{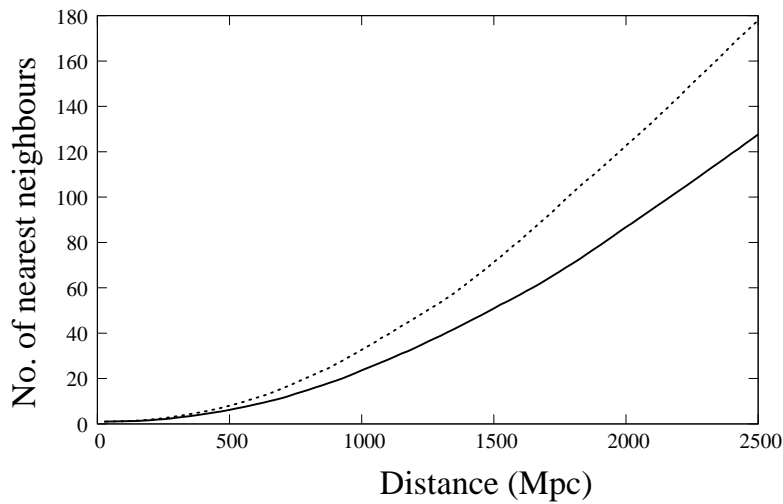}
    \caption{The number of nearest neighbours as a function of distance. The
upper curve (dashed) corresponds to angular diameter distance assuming a vacuum
dominated universe. The lower curve (solid) is obtained assuming comoving distance
in a matter dominated universe.
}   
\label{fig:nn}}
\end{figure}

\begin{figure}[!t]
  \centering{
    \includegraphics[width=4.5in,angle=0]{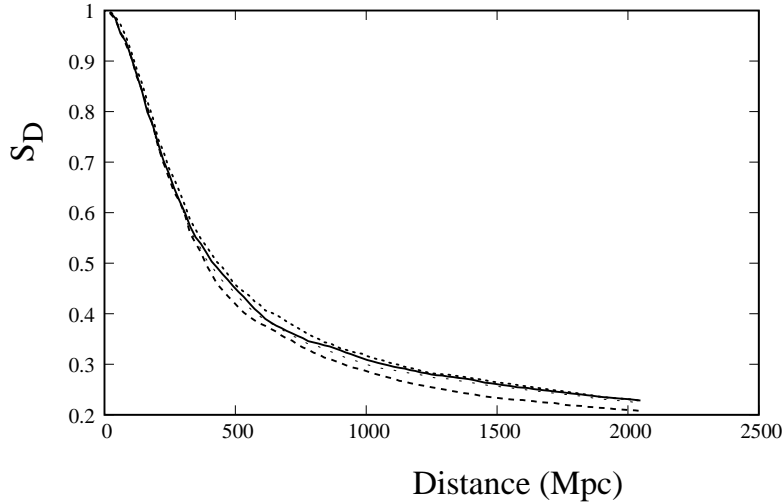}
    \caption{The fluctuations in the statistic $S_D$ for different choices of 
the random seed. 
Here we have set $n_B=-2.37$, $r_G=8$ Mpc and the number of points on the
grid $N_G=1024$. 
}   
\label{fig:scatter}}
\end{figure}


\begin{figure}[!t]
  \centering{
    \includegraphics[width=4.5in,angle=0]{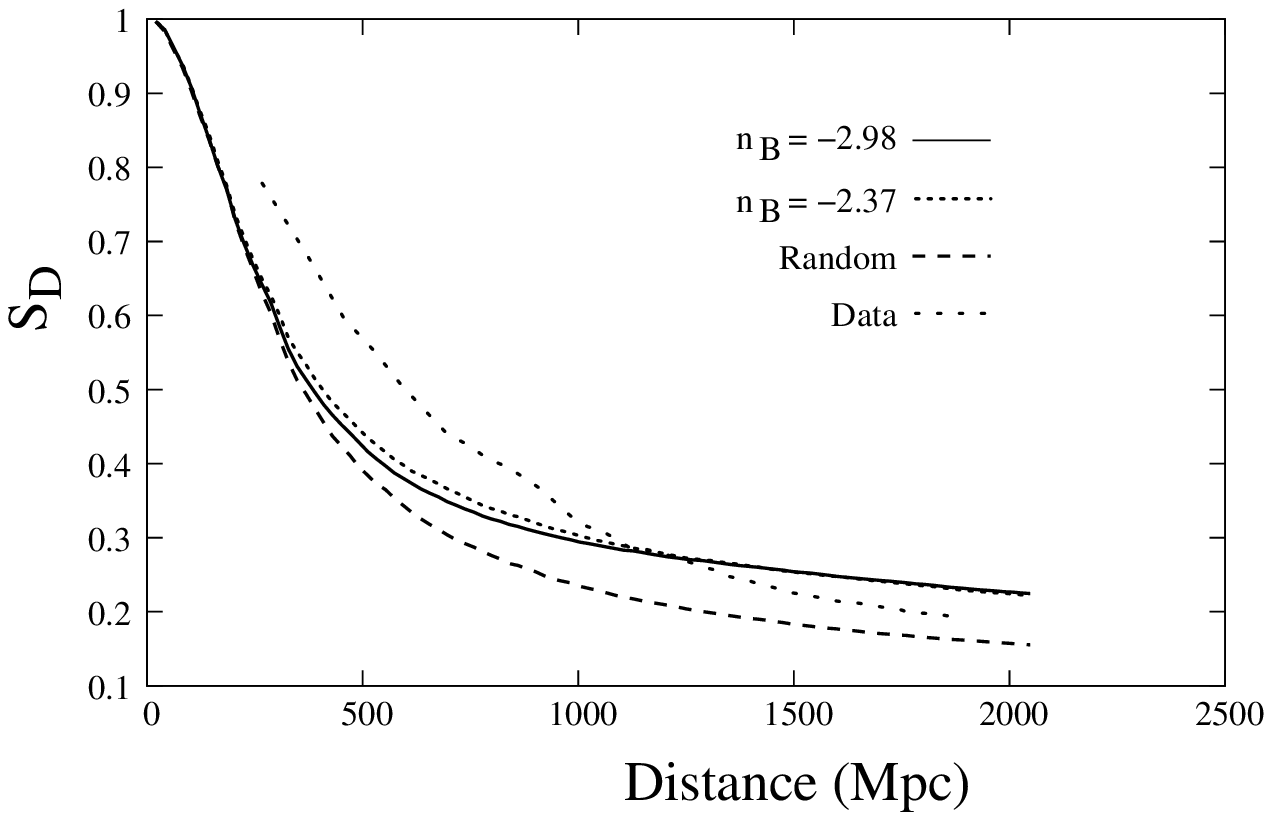}
    \caption{Comparison of the simulation results with observed data.
The dotted curve with large gaps shows the statistic obtained 
using observations. 
The remaining curves are obtained by simulations using parameters 
$n_B=-2.37$ (dotted with small gaps),
$n_B=-2.98$ (solid), and a completely random magnetic field (dashed).
Here we have set $r_G=8$ Mpc.}   
\label{fig:compare}}
\end{figure}

\section{Results}
\label{sc:results}
In this section we present the results of our numerical analysis. 
We first consider the case of flat space-time, ignoring cosmological
evolution. Next we give results for the case of expanding universe.  
We set the following parameter values: plasma density $n_e= 10^{-8}$
cm$^{-3}$,  frequency $\nu=10^6$ GHz,
coupling $g_\phi = 6.4\times 10^{-11}$ GeV$^{-1}$ and mass $m_\phi=0$. 
We shall use the spectral index $n_B= -2.37$ but will also explore
a range of values of this parameter.
We perform most of our simulations
on a grid size of $1024\times 1024 \times 1024$, with the observer located at the
center of the grid. The grid size is determined by the computing power 
available to us. 

\subsection{Flat space-time}
Here we present our results ignoring cosmological evolution.
We assume 200 quasars distributed randomly in space. 
The resulting statistic $S_D$ as a function of the distance
among sources 
is shown in Fig. \ref{fig:Sd}. Results are presented for four 
different values of the domain size $r_G=4, 8, 16, 32$ Mpc. The 
number of points on the grid are suitably scaled so as to keep the total
volume fixed. In Fig. \ref{fig:Sd} the statistic is computed 
by including all the sources
that lie within a particular distance from a source. 
This fixes the number
of nearest neighbours to be included for any distance. The real
data has a total of 355 sources. The relationship
between the number of nearest neighbours and the distance for the real data is shown in 
Fig. \ref{fig:nn}. We find that for a distance of 25 Mpc, the number of nearest neighbours is
approximately 1. It becomes close to 2 only at a distance of about 200 Mpc.   
In Fig. \ref{fig:scatter} we show the fluctuations in the statistic $S_D$ for a 
given choice of parameters $n_B=-2.37$, $r_G=8$ Mpc and the number of points
on the grid, $N_G=1024$. Results are shown for four different seeds used 
to initialize the simulations. We see significant fluctuations and hence
to compare with data it is better to take the mean over several
different computations. 
Finally in Fig. \ref{fig:compare} we compare the simulation results with 
observations. The simulation results are shown for a mean of five different
initializing seeds choosing $n_B=-2.37$ and $-2.98$. Here we have set $r_G=
8$ Mpc and the number of points on the grid equal to 1024. 
We also show results for the case where the background 
magnetic field is taken to be completely random. 
We find that the results for correlated magnetic fields show reasonable
agreement with data. However, those with a random magnetic field are systematically
below observations. 
We also find a relatively mild dependence 
on the exponent $n_B$.
We conclude that pseudoscalar-photon mixing
in a background magnetic field, described by 
Eqs. \ref{eq:IFT} and \ref{eq:corr}, 
gives reasonable agreement with observations. 
The small deviations from the data can be easily adjusted
by fine tuning the parameter values.

\subsection{Expanding Universe}
In this section we present our results for the case of an expanding universe. 
We consider 355 sources, to exactly match the number of sources observed in the real data. The positions of these sources are also taken to be the same
as in the data. The medium parameters, i.e. the magnetic field strength, 
the plasma density are taken to be the same as 
in the earlier non-expanding universe. Here these values are assumed to 
be their comoving values. The observed frequency $\omega$ of the electromagnetic
wave is assumed to be 2 eV. The comoving domain size is taken to be 15 Mpc. 
The Hubble constant is set equal to 
$2.133 h\times 10^{-42}$ GeV, with $h=0.7$. Furthermore we assume matter
dominated universe. As we shall
see our results for the expanding universe are comparable to those obtained
for flat space-time and hence such details do not make a very large difference 
to our final results.  

\begin{figure} 
\begin{center}
\includegraphics*[angle=0, width=1.0\textwidth,clip]{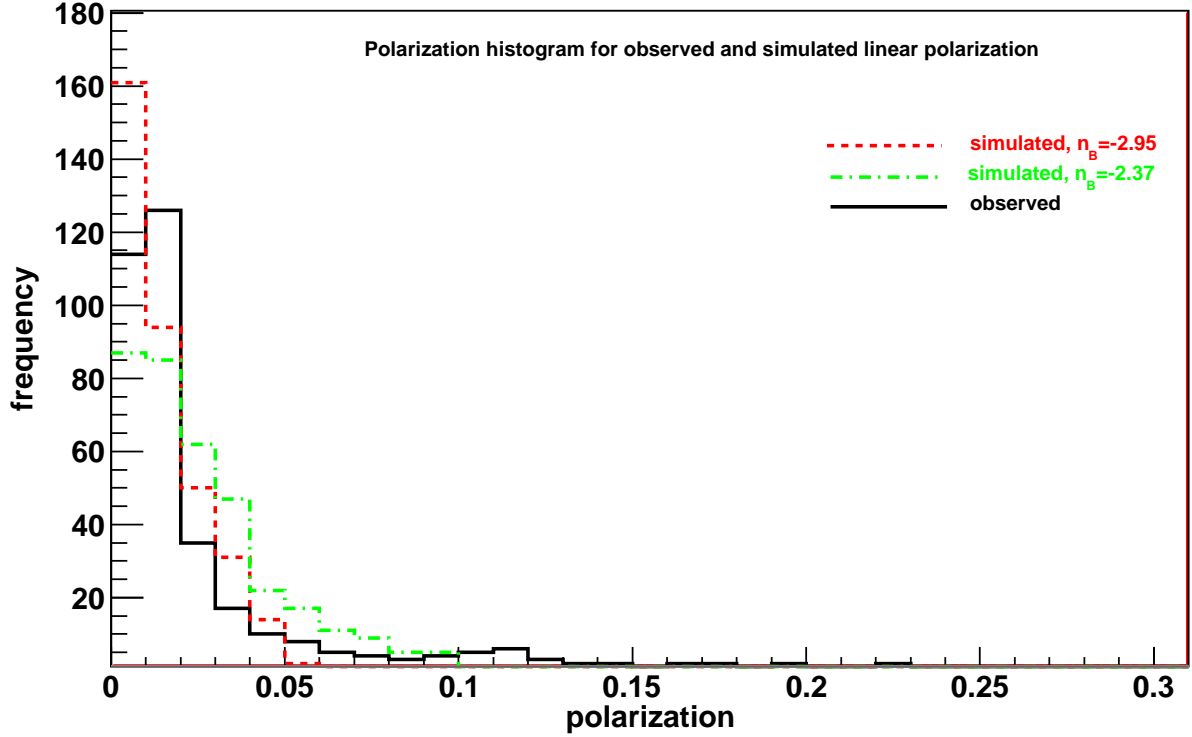}
\caption{The polarization histograms for the observed 355 sources (solid black curve) and for our simulations in an expanding universe. The red dashed curve
and the green dash-dotted curve show 
results for $n_B = -2.95$ and $n_B=-2.37$ respectively.}
\label{fig:expanding_hist}
\end{center}
\end{figure}

We test our model against the data of the 355 sources observed by \huts et al.
\cite{Hutsemekers:1998,Hutsemekers:2000fv,Hutsemekers:2005iz}.
On placing the sources in our simulations at the same positions as in the data, we perform the propagation and calculate the Stokes 
parameters. We calculate the coordinate invariant statistics
$S_D$, given in Eq. \ref{eq:statistic}, for the observed polarizations
and for the polarizations obtained from our simulations.
The distribution of the linear polarization both for observed data
and simulations is shown 
in Fig. \ref{fig:expanding_hist}. For the simulations we show results
for $n_B=-2.37$ and $-2.95$.

\begin{figure} 
\begin{center}
\includegraphics*[angle=0, width=1.0\textwidth,clip]{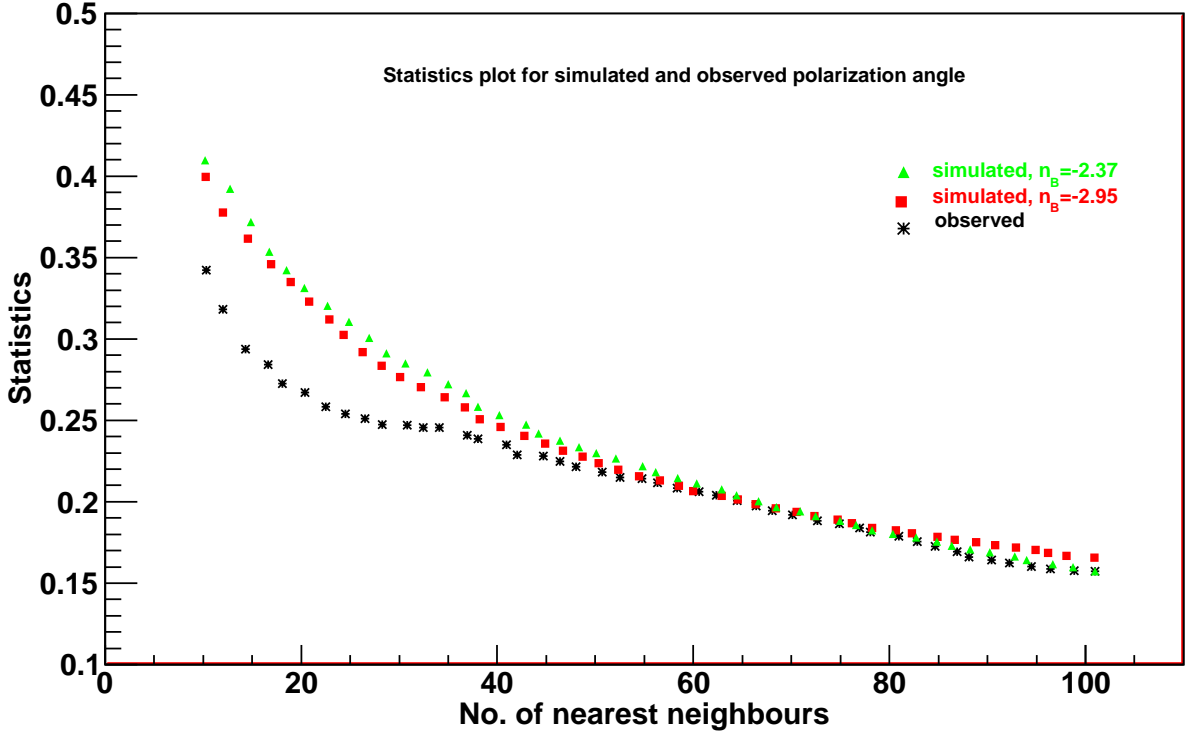}
\caption{The coordinate invariant statistic $S_D$ as a function of the
number of nearest neighbours for the 355 sources. The simulated results are
shown by green triangles ($n_B=-2.37$) and red squares ($n_B=-2.95$), while the black crosses represent the observed data.}
\label{fig:SD_expanding}
\end{center}
\end{figure}

In Fig. \ref{fig:SD_expanding}, we show the statistic $S_D$ for the data as well as the simulations. 
The simulation results are shown for a mean of five computations.
We see that for both $n_B=-2.37$ and $-2.95$ the simulations show
reasonable agreement with data.   
As mentioned above here we have set the domain size $r_G = 15$ Mpc. It might
be more reasonable to choose a smaller domain size. 
We find that as we reduce the domain size to $r_G = 12$ Mpc,  
our statistic, $S_D$, becomes significantly 
larger than data. Hence we find even
larger correlations in the optical polarizations. It may be possible
to play with parameters to make our simulation results agree with 
observations in this case also. For example, this may be accomplished 
by decreasing the background magnetic field and/or the 
pseudoscalar-photon coupling constant $g_\phi$.
However we postpone a detailed fit to future work since it requires 
extensive numerical simulations. 

\begin{figure} 
\begin{center}
\includegraphics*[angle=0, width=1.0\textwidth,clip]{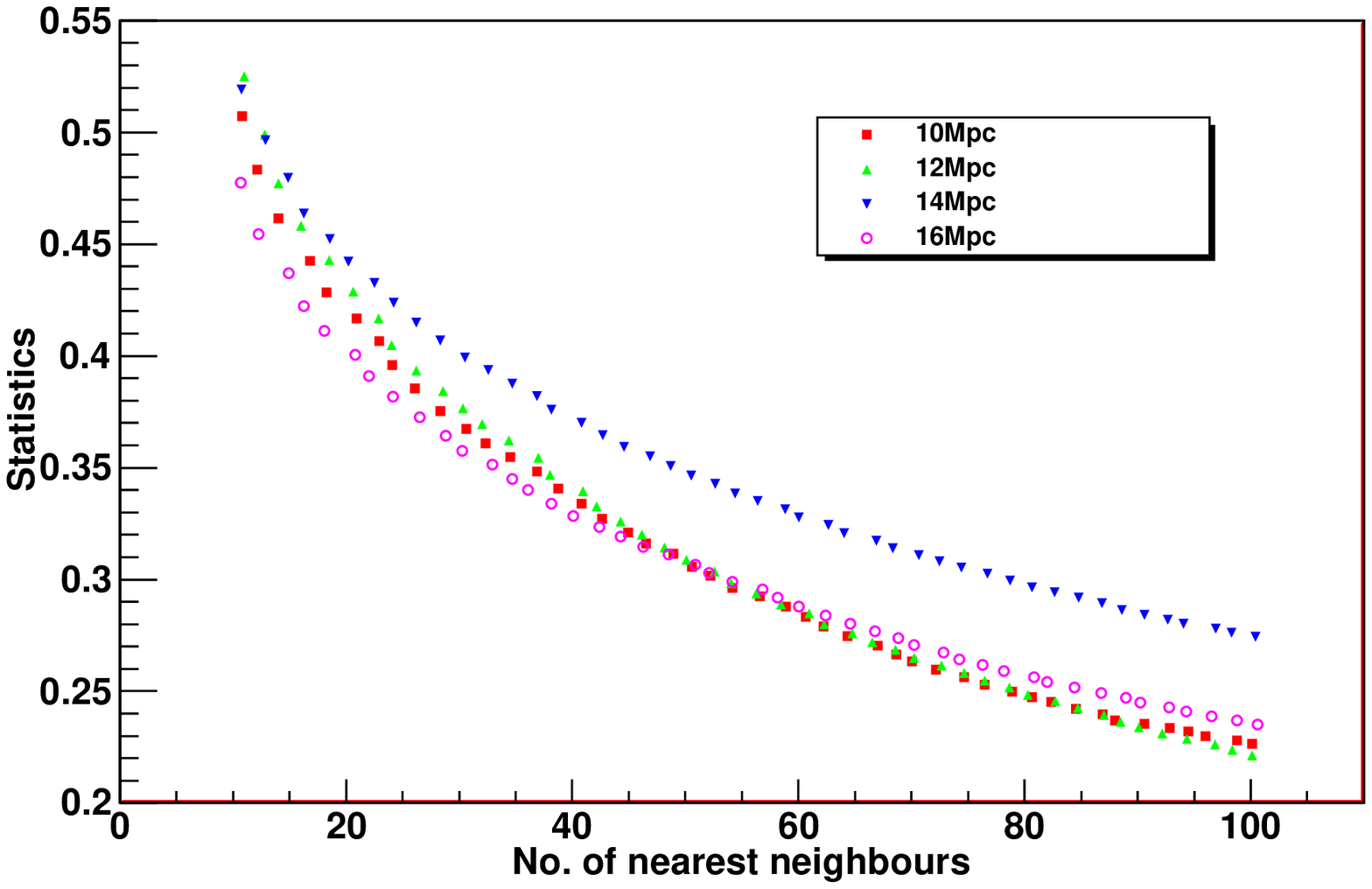}
\caption{The coordinate invariant statistic $S_D$ as a function of the
number of nearest neighbours for 355 sources. The sources are located
at the same angular positions as real sources but their radial distance
is reduced so that we can use a smaller grid of size $256\times 256\times
256$. Results are shown for four different domain sizes $r_G=$ 10 Mpc (red squares), 12 Mpc (green triangles), 14 Mpc (blue inverted triangles) and 16
Mpc (purple circles), after averaging over 40 different simulations.
Here we have set $n_B=-2.37$.}
\label{fig:SD_256}
\end{center}
\end{figure}

We may get some idea about the dependence of the statistic, $S_D$, on the
domain size by performing simulations on a smaller grid. For this purpose
we choose a grid size of $256\times 256\times 256$. We again take all the
355 sources but place them at a smaller radial distance with their 
polarizations and angular positions taken to be same as that of real data. 
The resulting statistic, $S_D$, as a function of the number of nearest
neighbours is shown in Fig. \ref{fig:SD_256} for four different domain sizes.
The results are shown after averaging over 40 different simulations in order
to reduce the effect of fluctuations. We find that $S_D$ shows an oscillatory
behaviour as a function of the domain size. It increases as we increase 
the domain size from 10 Mpc to 14 Mpc. However with a further increase
from 14 Mpc to 16 Mpc it starts to decrease. 
Similar trend is seen for a larger grid of $512\times 512\times 512$. 
This suggests that for the range of parameters we consider, $S_D$ shows
an oscillatory dependence on the domain size. Furthermore the variation is
significant but not very large. These results suggest   
that for the distance scale corresponding to the real data, 
the results may also show oscillations. Hence we expect to find a suitable 
fit to data even for smaller domain sizes with other parameters similar to
what were chosen for the fit shown in
Fig. \ref{fig:SD_expanding}. 

\section{Mixing in a more general Background}
As mentioned in the introduction a potential problem with our hypothesis
is that the observed circular polarization in quasars is found to be 
negligible \cite{Hutsemekers:2005iz,Hutsemekers:2010}. 
However our model predicts a
large circular polarization. We argued in the introduction that our 
model of the background medium, i.e. the magnetic field and the plasma
density, may be unrealistic. The medium is expected to show fluctuations
in time and space at any length scale relevant to propagation over 
cosmological distances. Here we have   
assumed that the magnetic field and the plasma density are time independent.
Furthermore we have assumed that the 
the magnetic field is uniform in a particular domain. In this section 
we determine
how these assumptions affect our predictions for circular polarization.

\subsection{Space dependent magnetic field}
We first examine how small scale spatial fluctuations in the magnetic field
affect final results. We 
will continue to assume a uniform plasma density throughout this section.
As discussed above,
the magnetic field is likely to have fluctuations even on the scale
of a single domain. Here we assume a simple model of space varying magnetic 
field, given by, 
\beq {\bf B} = \mathcal{B}_0(\cos^2 \alpha ~ \hat{\bf x}+ \sin^2 \alpha ~\hat{\bf y}) \eeq
where, \beq \alpha = \frac{2 \pi}{\lambda} \times z \eeq 
and $\lambda$ is the wavelength for the $B$ field fluctuations.
The rate at which the magnetic field changes with position $z$ can be
varied by changing the wavelength. Furthermore we point out that the
field has been chosen such that its mean is not zero.
   The $A$ and $\chi$ field equations with this background can be written 
as, 
\beq
(\omega+ \mathrm{i}\partial_{z})          \left(\begin{array}{c}
 {\cal A}_1\\ {\cal A}_2 \\ \chi            \end{array} \right)   -
M \left( \begin{array}{c} {\cal A}_1\\ {\cal A}_2 \\ \chi    \end{array} \right) =0, \eeq
where,
\beq M = \left(\begin{array}{ccc}
        
 \frac{\omega^{2}_{p}}{2\omega}     &    ~~~ 0     & ~~~   -\frac{g_{\phi}}{2} \mathcal{B}_0 \cos^{2} \alpha \\
  0 ~~~    & \frac{\omega^{2}_{p}}{2\omega}     &~~~   -\frac{g_{\phi}}{2}     \mathcal{B}_0 \sin^{2} \alpha \\
  -\frac{g_{\phi}}{2} \mathcal{B}_0 \cos^{2} \alpha  & ~~~ -\frac{g_{\phi}}{2} \mathcal{B}_0 \sin^{2} \alpha  & ~~~ \frac{m^{2}_{\phi}}{2\omega} \end{array}\right). \eeq
Here, we have approximated $(\omega^{2} +\partial^{2}_{z}) \approx 2\omega (\omega + \mathrm{i} \partial_{z})$ \cite{Raffelt:1987im} 
and  we have taken  $\bf{A}$ as $( {\cal A}_1 ~\hat{\bf x} + {\cal A}_2 ~\hat{\bf y})$. These equations are same as those given in Section [\ref{sc:solution}]
with the scale factor $a=1$.   

Here we do not assume that the background magnetic field is changing 
sufficiently slowly so that we can choose basis vectors such that one
of them points parallel to the transverse background magnetic field at any
point $z$ along the path of the wave. 
This approximation would be valid if $\lambda$ is very large but would
break down for small $\lambda$. Instead we 
directly solve these equations numerically in terms of   
the density matrix  which is defined as, 
\beq \rho(z)  = \left (\begin{array}{ccc}
<{\cal A}^{\ast}_1(z) {\cal A}_1(z) >  &\tab  <{\cal A}^{\ast}_1(z) {\cal A}_2(z) >  &\tab  <{\cal A}^{\ast}_1(z) \chi(z) > \\ 
<{\cal A}^{\ast}_2(z) {\cal A}_1(z) >  &\tab  <{\cal A}^{\ast}_2(z) {\cal A}_2(z) >  &\tab  <{\cal A}^{\ast}_2(z) \chi(z) > \\ 
<  \chi^{\ast}(z)     {\cal A}_1(z) >  &\tab  <   \chi^{\ast}(z)    {\cal A}_2(z) >  &\tab  <   \chi^{\ast}(z)    \chi(z)> \\ 
\end{array}\right). \eeq
The Stoke's parameters can be written as, 
\bea 
I(z)= <{\cal A}^{\ast}_1(z) {\cal A}_1(z) > +  <{\cal A}^{\ast}_2(z) {\cal A}_2(z) >\\
Q(z)= <{\cal A}^{\ast}_1(z) {\cal A}_1(z) > -  <{\cal A}^{\ast}_2(z) {\cal A}_2(z)>\\
U(z)= <{\cal A}^{\ast}_1(z) {\cal A}_2(z) > +  <{\cal A}^{\ast}_2(z) {\cal A}_1(z) >\\
V(z)= \mathrm{i}( <{\cal A}^{\ast}_1(z) {\cal A}_2(z) >  - {\cal A}^{\ast}_2(z) {\cal A}_1(z) >)   
\eea

\begin{figure}
\begin{center}
\includegraphics*[angle=0, width=1.0\textwidth,clip]{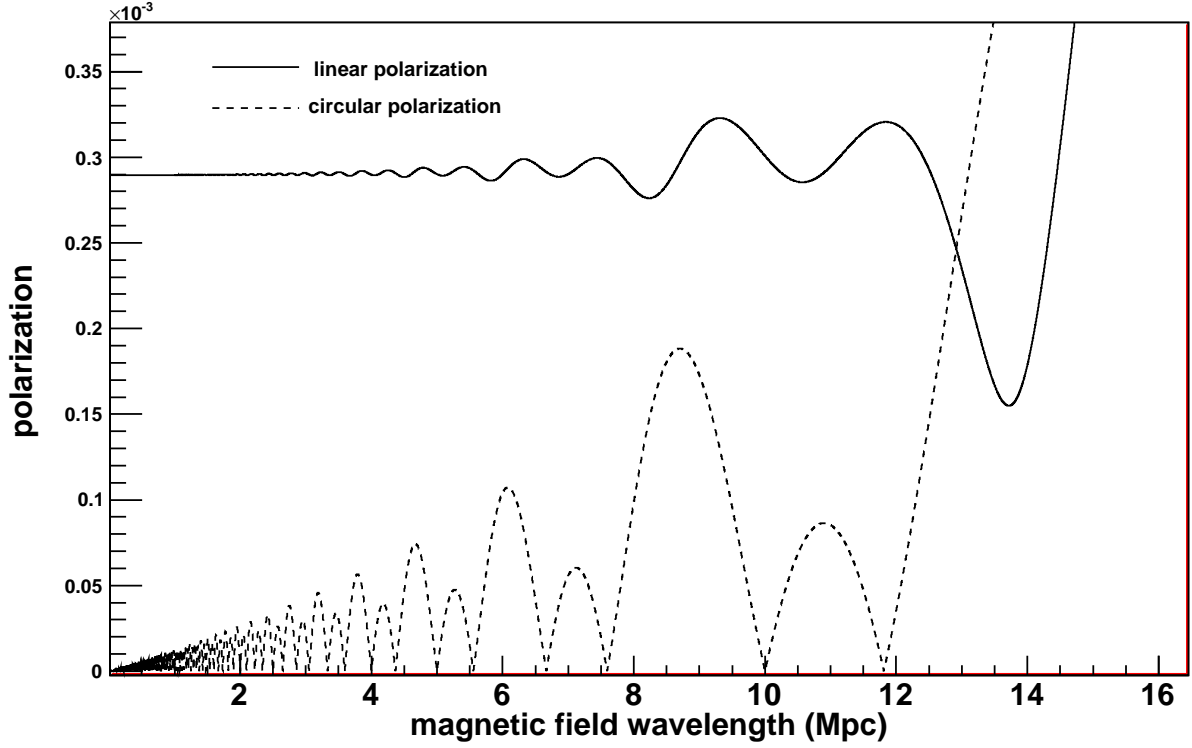}
\caption{The variation of linear (solid curve) and circular polarization
(dashed curve) with respect to 
the wavelength $\lambda$ of the
background magnetic 
field. }
\label{fig:Bfield}
\end{center}
\end{figure}

We assume that the wave is initially unpolarized and propagate 
in this medium over a distance of 10 Mpc. 
The resulting circular polarization $|V|/I$ and the linear polarization
$\sqrt{Q^2+U^2}/I$ is shown 
 in Fig. \ref{fig:Bfield} as a function of the wavelength $\lambda$ of
the background magnetic field. The figure clearly shows that for small 
$\lambda$, which corresponds to 
rapid fluctuations, the circular polarization is   
almost negligible as compared to the linear polarization. 
The value of linear polarization for small $\lambda$
is comparable to that at large $\lambda$, where it starts
to show large fluctuations.  
The circular
polarization becomes comparable to linear polarization if the background
varies slowly. 
Hence we clearly see that circular polarization is 
much smaller than linear polarization if the background magnetic field
shows spatial fluctuations on sufficiently small length scales. 

Here we have presented results only for a single domain. We do not attempt
the more ambitious task of integrating over all the domains. That would  
be much more computationally intensive in comparison to the calculation performed
in this paper since it would involve solving a system of differential equations
in each domain. Furthermore it would depend on the precise model we use for
the magnetic field in each domain. In any case our analysis in this section
clearly shows that there exist very reasonable model of magnetic field which would
lead to highly suppressed circular polarization. 

\subsection{Time dependent background magnetic field}
Another important issue that we consider is the possible time
dependence of the background magnetic field.
If the background magnetic field is time dependent, the pseudoscalar
produced due to mixing with photons will not have the same frequency as
that of the incident photon. Hence its correlation with photons may be 
significantly reduced. The circular polarization arises primarily due
to reconversion of pseudoscalars back into photons. This generates a  
relative phase among the different components of the electromagnetic wave
and hence leads to circular polarization. We see this explicitly by
considering the evolution of the density matrix in a particular domain,
given in Ref. \cite{Agarwal:2008ac}, reproduced here for convenience,
\begin{eqnarray}
<A_{||}(z) A_{||}^{*}(z)> & = & \frac{1}{2} <A_{||}(0) A_{||}^{*}(0)>
 \Bigl[ 1 + \textrm{cos}^{2} \; 2\theta + \textrm{sin}^{2} \; 2\theta \;
       \textrm{cos}[z(\Delta_{\phi} - \Delta_{A})] \Bigr]  \nonumber \\
        & + & \frac{1}{2} <\phi(0) \phi^{*}(0)>
\Bigl[ \textrm{sin}^{2} \; 2\theta - \textrm{sin}^{2} \; 2\theta \; \textrm{cos}[z(\Delta_{\phi} - \Delta_{A})]   \Bigr]  \nonumber \\
        & + & \Biggl\{ \frac{1}{2} <\phi(0) A_{||}^{*}(0)>
              \Bigl[ \textrm{sin} \; 2\theta \; \textrm{cos} \; 2\theta - \textrm{sin} \; 2\theta \; \textrm{cos} \; 2\theta                                                 \; \textrm{cos}[z(\Delta_{\phi} - \Delta_{A})] \nonumber \\
        & - &        \textrm{i} \; \textrm{sin} \; 2\theta \; \textrm{sin}[z(\Delta_{\phi} - \Delta_{A})]
     \Bigr] + \textrm{c.c.}
              \Biggr\}  \\                                              
        <A_{\bot}(z) A_{\bot}^{*}(z)> & = & <A_{\bot}(0) A_{\bot}^{*}(0)>  \\
<A_{\bot}(z) A_{||}^{*}(z)>   & = & <A_{\bot}(0) A_{||}^{*}(0)>        
        \Bigl[ \textrm{cos}^{2} \; \theta \; \textrm{e}^{\textrm{\scriptsize i}\;Fz} + \textrm{sin}^{2} \; \theta \;                                                   \textrm{e}^{\textrm{\scriptsize i}\;Gz}
        \Bigr]  \nonumber \\
        & + & <A_{\bot}(0) \phi^{*}(0)>        
        \Bigl[ \textrm{sin} \; \theta \; \textrm{cos} \; \theta
    \left( \textrm{e}^{\textrm{\scriptsize i}\;Fz} - \textrm{e}^{\textrm{\scriptsize i}\;Gz}
      \right)\Bigr]  
\end{eqnarray}
Here the vector ${\bf A}={\bf E}/\omega$, $\omega$ is the frequency of the
electromagnetic wave, $\phi$ is the pseudoscalar field and 
$A_\parallel$ and $A_\perp$ respectively represent
the components of ${\bf A}$, parallel and perpendicular to the
transverse component of the background
magnetic field. These equations give the density matrix elements after
propagating through one domain, assuming a space and time independent 
magnetic field and plasma
density. The circular polarization is governed by the correlation 
$<A_\perp(z)A^*_\parallel(z)>$. Let us assume that initially this correlation
is zero. Then after propagating through one domain, it becomes non-zero 
only if initially the correlation $<A_\perp(0)\phi^*(0)>$ is non-zero.
We have assumed that at the source all the cross correlations are zero, 
i.e. the beam is unpolarized and the pseudoscalar field is zero. Hence
circular polarization can be produced only if, after propagating through 
some domains, the correlation $<A_\perp\phi^*>$ becomes non-zero. 
However here we argue that if the background magnetic field is time-dependent,
the frequency of the pseudoscalar field produced due to mixing would be
different from that of the electromagnetic field. Hence its correlation would
be significantly reduced. 

The pseudoscalar field equation may be written as,
\begin{equation}
        {\partial^2 \phi\over \partial t^2} - \nabla^2\phi + m_\phi^2\phi =  g_{\phi}{\bf E} \cdot \boldsymbol {\mathcal B}\ . 
\label{pseudoscalar}
\end{equation}
where $\boldsymbol{\mathcal B}$ represents the background magnetic field. Let us assume
that the background magnetic field also undergoes fluctuations in time. 
Hence we may make a Fourier decomposition of this field. Here we assume
that it contains only one significant component with frequency
$\Omega$, i.e. $|\boldsymbol{\mathcal B}| \sim \cos(\Omega t)$. 
The frequency of the electric field is $\omega$. It is clear from Eq. 
\ref{pseudoscalar} that the frequency of the pseudoscalar produced 
has to be $\omega'=\omega\pm\Omega$. 
 
Let us assume that the incident beam has a narrow spectral distribution, 
specified by the function $A(\omega)$, which denotes one of the components
of the electromagnetic wave. We require correlations such as, 
$<A(\omega')\phi^*(\omega')>$, where $\omega'=\omega\pm \Omega$. Since
$\phi(\omega')\propto A(\omega)e^{-i\omega't}$, we have
\be
<A(\omega')\phi^*(\omega')>\  \propto\  <A(\omega')A^*(\omega)>
\ee
It is clear that if $\omega'$ is sufficiently different from $\omega$, this
correlation would be vanishingly small. We expect this correlation
to decrease rapidly with increase in the difference between the two frequencies.
We point out that the background medium is expected
to have fluctuations on all time scales. It is only after averaging out 
short time fluctuations that we expect it to show a somewhat smooth behaviour.
Hence it may not be reasonable to assume that $\Omega$ is close to zero.
We point out that, although these temporal fluctuations may significantly 
affect the circular polarization, the mixing phenomenon would still generate
significant linear polarization, as long as the mean value of the transverse
component of the magnetic field is significantly different from zero. 
This is because the electromagnetic wave component parallel to the 
transverse 
magnetic field will decay into pseudoscalars and the perpendicular component
will not. Hence we find that circular polarization may be significantly
reduced compared to linear polarization if the background magnetic field shows
sufficiently rapid fluctuations in time.  

\section{Conclusions}
We have shown by direct simulations, both in a static and expanding universe, 
that the hypothesis of pseudoscalar-photon
mixing is able to explain the observed alignment of optical polarizations
from distant quasars for standard parameter values of the intergalactic
magnetic field and plasma density. The value of pseudoscalar-photon 
coupling is taken to be close to its observational limit. The simulation
results show reasonable agreement with data 
both with the distribution of linear polarization
as well as the coordinate invariant statistic. 
Given the uncertainties in the medium parameters we have not performed the 
exercise of finding the best fit to the data. In any case before attempting
a detailed fit it is necessary to perform further calculations to 
conclusively establish that pseudoscalar-photon mixing is indeed responsible
for the alignment. One potential problem with this explanation, as discussed
in the introduction, is the relatively large circular polarization predicted by 
this phenomenon \cite{Hutsemekers:2011}. However, there are many sources
of decoherence which need to be properly taken into account before 
making a firm conclusion on the strength of circular polarization.
We have explicitly shown that if the magnetic field shows  
fluctuations over the distance scale of individual domains, the circular
polarization predicted by this phenomenon is significantly reduced in 
comparison to linear polarization. We find this phenomenon even for a 
relatively large wavelength
of fluctuations, of order 1 Mpc, of the background magnetic field.
Furthermore 
the agreement we find with data is potentially 
encouraging and hence this proposal
may be taken seriously. It is possible that even if this precise mechanism
is found not to explain the data consistently, a suitable generalization
might work. We postpone such issues for future research.



\section*{Acknowledgements}
We thank Archana Kamal for useful discussions. We thank Subhayan Mandal for
sharing with us his unpublished work on axion-photon mixing 
in an expanding universe.

\bibliographystyle{apsrev}
\bibliography{Optical}
\end{document}